\documentclass[aps,prl,twocolumn,superscriptaddress,amsmath,amssymb,showpacs]{revtex4}
\usepackage[dvips]{graphicx}
\begin{document}

\title{Guiding and Trapping of Electron Spin Waves in Atomic Hydrogen Gas}

\author{O. Vainio}
\author{J. Ahokas}
\author{S. Novotny}
\author{S. Sheludyakov}
\affiliation {Department of Physics and Astronomy, University of Turku, 20014 Turku, Finland}
\author{D. Zvezdov}
\affiliation {Department of Physics and Astronomy, University of Turku, 20014 Turku, Finland}
\affiliation {Kazan Federal University, 420008, 18 Kremlyovskaya St, Kazan, Russia.}
\author{K.-A. Suominen}
\author{S. Vasiliev}
\email{servas@utu.fi}
\affiliation {Department of Physics and Astronomy, University of Turku, 20014 Turku, Finland}

\date{\today}

\begin{abstract}
We present a high magnetic field study of electron spin waves in atomic hydrogen gas compressed to high densities of $\sim 10^{18}$ cm$^{-3}$ at temperatures ranging from 0.26 to 0.6 K. We observed a variety of spin wave modes caused by the identical spin rotation effect with strong dependence on the spatial profile of the polarizing magnetic field. We demonstrate confinement of these modes in regions of strong magnetic
field and manipulate their spatial distribution by changing the position of the field maximum.
\end{abstract}


\maketitle

%\textbf{Introduction} Definition of a quantum gas. Magnons in QG. Comparison with ferromagnets.

Spin and magnetization oscillations are well known phenomena in condensed matter physics.  In solids wave-like spin excitations are caused by long-range dipolar or exchange interactions \cite{Gurevich96}. Quantized spin excitations (magnons) may exhibit collective quantum
phenomena similar to BEC and superfluidity in quantum gases and liquid He. Observation of these effects in superfluid $^3$He \cite{Bunkov07}
and ferromagnets \cite{Demokritov06} have been extensively discussed recently. In gases spin waves may also propagate although interactions are weak and occur during short collisional events. Quantum gases represent a special case where de Broglie wavelength $\Lambda_{th}$ exceeds the characteristic range $a$ of the interatomic potential and therefore the identical-particle exchange
effects in atomic collisions appear long before the electrons of the atoms start to interact.
If two colliding atoms have different spin directions, exchange interaction leads to the rotation of both spins around their sum, a
phenomenon called the Identical Spin Rotation (ISR) effect \cite{Bashkin81,Laloe82}. Being accumulated in numerous collisions ISR leads to the
propagation of spin excitations. The efficiency of ISR is determined by the parameter $\mu \sim \Lambda_{th}/a$, which is large in the quantum
gas limit. Therefore, for cold gases there exists a range of densities $n$ and temperatures where $n^{-1/3}\gg \Lambda_{th}\gg a$, i.e. the gas is
in the quantum regime but not yet degenerate. This range is especially wide for smallest atoms like H ($a \approx 0.07$ nm). Experimentally
the ISR effect was first observed in the nuclear spins of electron-spin polarized H (H$\downarrow$) \cite{Johnson84}, then in $^3$He gas
\cite{Nacher84} and recently also in ultracold $^{87}$Rb vapors \cite{McGuirk02,Deutsch10}. ISR waves were predicted for electron spins of
H$\downarrow$ \cite{Bashkin81,Lhuillier85} but have not been observed so far. In this letter we report a quantitative study of electron
spin waves in H$\downarrow$ and verify that they are associated with ISR.

Spin transport in quantum gases is described by  the complex diffusion equation for the transversal spin-polarization $S_{+}=S_{x}+i S_{y}$ \cite{Levy84},
simplified for $S_{+}\ll S_{z}$ and $\mu\gg1$ to:
\begin{equation}
i \frac{\partial S_{+}}{\partial t}= D_0 \frac {\varepsilon}{\mu} \nabla^{2} S_{+} + \gamma \delta B_0 S_{+}, \label{ISR equation}
\end{equation}
where $D_0$ is the spin diffusion coefficient in unpolarized gas, $\varepsilon=+1$ for  bosons and $-1$ for fermions, $\gamma$ is the gyromagnetic ratio,
and $\delta B_0$ is the deviation of magnetic field from its average value $B_0$. Eq.~(\ref{ISR equation}) is similar to the Schr\"{o}dinger equation for a
particle with an effective mass $M^{*}=-\hbar \mu / 2D_{0} \varepsilon$ and potential energy given by the second term on the right-hand side. The behavior of such quasiparticles in an inhomogeneous magnetic field depends on the signs of the parameters $\mu$,
$\varepsilon$, and $\gamma$. The spin-wave quality factor $\mu$ depends on the details of the interatomic potential. For H $\mu < 0$, and for $^3$He it changes sign  from negative to positive when temperature is lowered below $\approx0.5$ K \cite{Lhuillier83}.
From Eq.~(\ref{ISR equation}) we see that similarly to low-field seeking atomic species the nuclear spin waves of H have lower potential energy in regions of smaller magnetic field. Due to $\gamma$ being negative the electron spin waves behave as high field seekers. Similar behavior is expected for $^3$He at $T>0.5$ K ($\mu>0$, $\varepsilon=-1$). In addition, as $\gamma_e /\gamma_n \approx 650$, the strength of the potential energy for electron spin waves is much larger than that for nuclei. Consequently local magnetic field maxima become effective potential wells. We will show in this letter that similarly to real particles ISR spin waves in H$\downarrow$ may be spatially confined and manipulated by magnetic forces.

\begin{figure}
\includegraphics[width=8 cm]{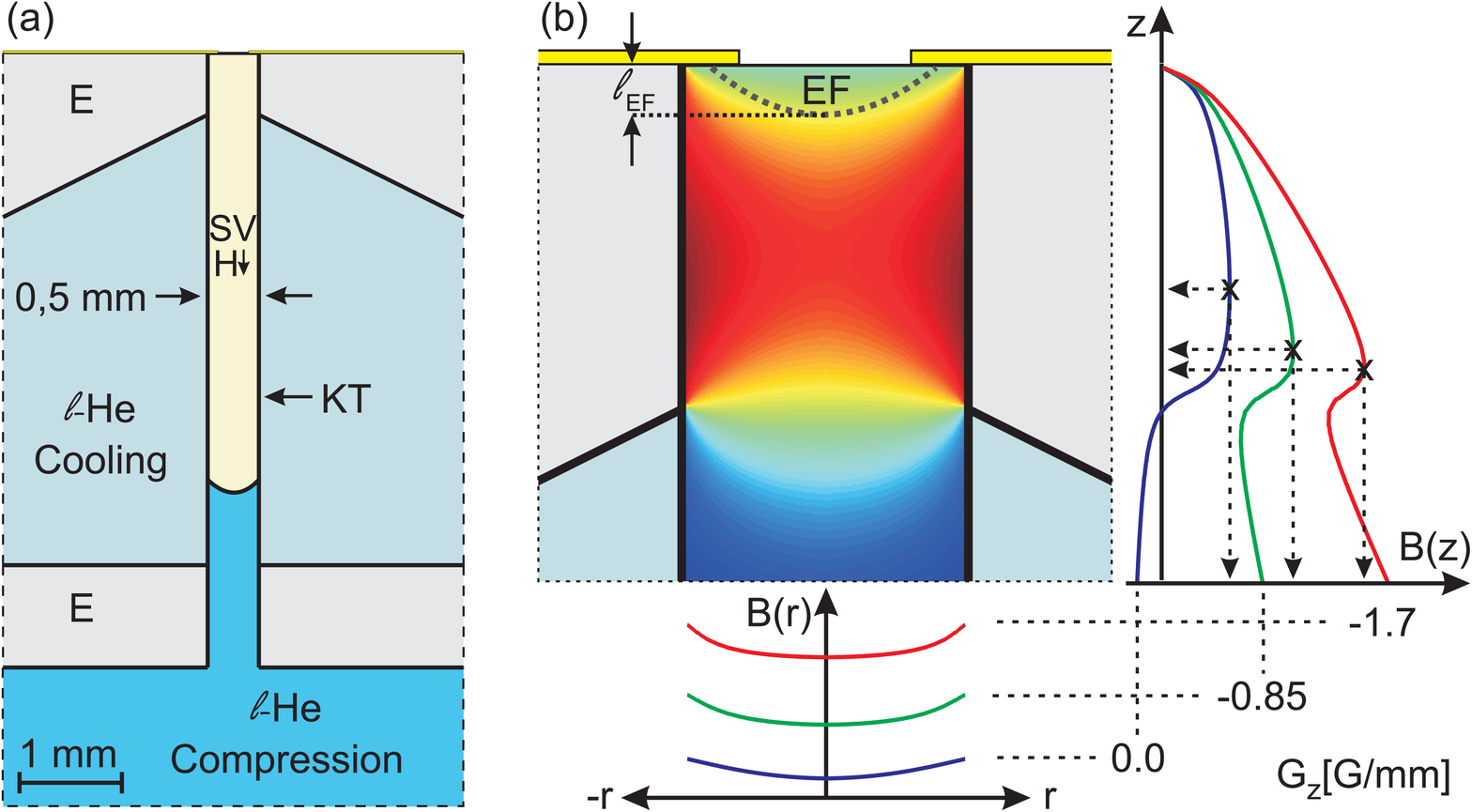}
\caption{(Color online) (a) Schematic drawing of the sample cell showing the sample volume (SV), Kapton tube (KT), liquid-helium volumes (l-He)
and epoxy structures (E). (b) Magnetic field inhomogeneity due to magnetized epoxy. The field amplitude in the top of SV is shown as
a colormap, black (red) corresponding to the strongest and blue (white) to the weakest field. Axial (z) and radial (r) cuts of the field
profile through the field maximum are also shown for three different additional linear gradient values: -1.7 (red/black), -0.85 (green/dark) and 0 (blue/light) G/mm.}
\label{cell}
\end{figure}

In our experiment H$\downarrow$ gas is hydraulically compressed up to $n\sim 10^{18}$ cm$^{-3}$, as described in \cite{Ahokas08}. The
sample is located in a thin-walled (2 $\mu$m) 0.5 mm diam. Kapton tube (KT) below the flat mirror of a Fabry-Perot resonator
(FPR) (see Fig. \ref{cell}). The sample volume (SV) is coupled to the rf. field of the resonator via an evanescent field (EF) beneath a subcritical (0.4
mm dia.) orifice in the FPR-mirror. The evanescent field amplitude $H_1(z)$ has an approximately Gaussian shape decreasing downwards, with an
effective height of $l_{EF}\approx80$ $\mu$m \cite{JarnoPhD}. The bottom of SV is limited by the meniscus of superfluid helium. The height
$L$ of SV depends on the position of the helium level and can be varied from 6.5 to 0.5 mm.
For smaller $L$ the sample evolves into a bubble. The Kapton tube is glued with Stycast 1266 to an epoxy disk with conical cross section. The
thickness of the disk at the tube wall is $\approx 0.5$ mm. The outer surface of KT below the epoxy disk is flushed with superfluid
helium cooled to 200-600 mK in a separate heat exchanger coupled to the mixing chamber of the dilution refrigerator. This construction
allows to reach high densities with linear response of the FPR to the  electron spin resonance (ESR) in H$\downarrow$. For the excitation and detection of spin waves
a highly inhomogeneous rf. field combined with a controllable static field are used. The homogeneity of the magnetic field in the compression region was limited to $4\pi M \sim 0.8$ G by the weakly paramagnetic epoxy disk, the field of which is equivalent to that of a miniature solenoid located at the tube wall, creating a saddle shaped field with a
maximum located near the wall (see Fig. \ref{cell}(b)). We were able to reduce the inhomogeneity to $\approx 20$ mG over the rf. field
region using a set of linear and parabolic shim coils, which were also capable of generating linear axial field gradients up to
$G_z=\pm 4$ G/mm.

\begin{figure}
\includegraphics[width=8 cm]{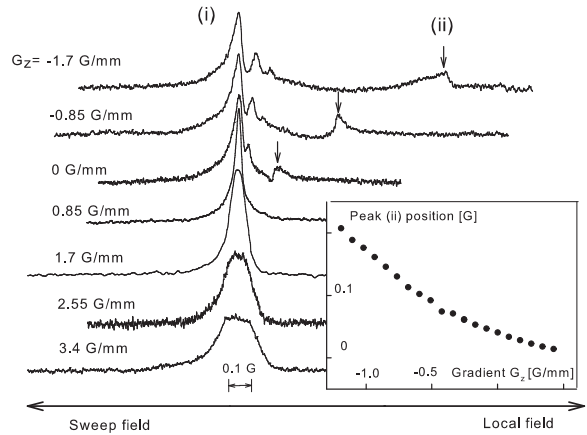}
\caption{ESR spectra for different values of axial magnetic field gradient. The resonance frequency is fixed and magnetic field is swept through
the resonance. The regions of the sample volume located in larger local fields will get in resonance at smaller sweep field. The direction
of the field axis is reversed to get the local field increasing from left to right. Insert: peak (ii) position as a function of $G_z$.} \label{Spectra in Grad}
\end{figure}

\begin{figure}
\includegraphics[width=8 cm]{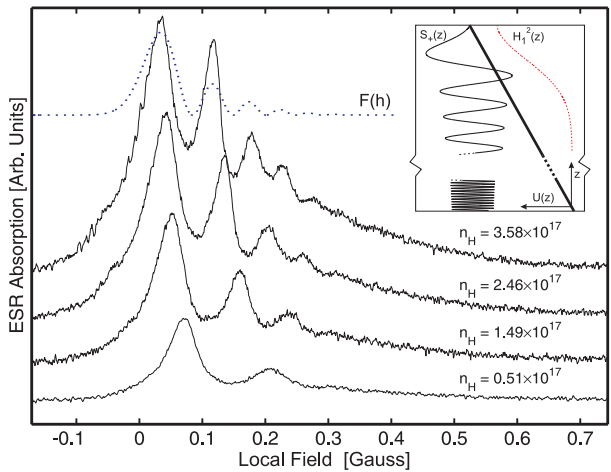}
\caption{ESR spectra observed for samples $G_{z}=-4$ G/mm. Dotted line: the simulated contribution $F(h)$ of spin waves to the ESR absorption. In the insert the orientations of the potential $U(z)$ (thick solid line), a spin
wave mode $S_{+}(z)$, and the evanescent field $H_{1}^{2}(z)$ are shown emphasizing the upper part of the sample volume. }
\label{NegGradSpectra}
\end{figure}

The H$\downarrow$ samples were studied by ESR technique at 128 GHz \cite{OurESR}. The spectra were recorded by applying a constant frequency from a highly stabilized mm-wave source while sweeping the static magnetic field offset $h$ through the resonance. We utilized the
\textit{b-c} transition, since the population of the hyperfine state \textit{a} is vanishingly small due to rapid exchange recombination (for the notations of the hyperfine states see,
e.g. \cite{BlueBible}). At small enough densities $n \lesssim 5 \times 10^{16}$ cm$^{-3}$ the ESR lines were inhomogeneously broadened due to the spurious field of the epoxy ring. A small positive gradient $G_{z} \approx 1$ G/mm best compensated the inhomogeneity
resulting in the smallest line widths. For $n > 10^{17}$ cm$^{-3}$ we observed two distinct changes in the shape of the
ESR lines: (i) the main absorption peak was
split into several narrow lines whose position and separation depended on $n$ and $G_z$, and (ii) within the range of
$G_z \approx -2...+0.85$ G/mm an extra peak appeared on the right side of the spectra. These features are summarized in Figs. \ref{Spectra in Grad} and \ref{NegGradSpectra}. The
separation of this peak (ii) from the main ESR line increased with $G_z$ and remained visible even at the lowest densities $n \sim 10^{16}$ cm$^{-3}$ at which we were still able to resolve ESR lines. However, the peak (ii) was never observed on the left from the main resonance
line not even with sufficiently large positive gradients. Neither was there any dependence of the peak positions on the height of the gas sample, except for
$L<0.5$ mm, when the sample evolved into a small bubble.

Next we consider the splitting of the main ESR peak for high densities and large negative $G_z$ (feature (i)). To estimate the expected separation between the ISR spin wave modes we assume an integer number of their half-wavelengths between the walls of the compression
region. From the ISR spin-wave dispersion law $\omega=\frac{D_0}{\mu}k^2$ we get $\Delta \omega\lesssim \pi^2 D_0 / \mu L^2$. Taking the
minimum sample height $L\approx 0.5$ mm and the values $D_{0}n= 1.5\times 10^{18}$ cm$^{-1}$s$^{-1}$ and $\mu\approx 7$ from experiments
with nuclear spin waves in H$\downarrow$ \cite{Johnson84}, we find that the spin wave modes will be separated by $\lesssim1.5$ kHz at $n=10^{17}$ cm$^{-3}$. This corresponds to $\lesssim0.5$ mG in the field sweep and cannot be resolved in the experiment. It is thus
unlikely that the peaks seen in Fig.~\ref{Spectra in Grad} and \ref{NegGradSpectra} correspond to individual spin wave modes, but rather to a number of overlapping modes.
It turns out that just in the case of strong negative field deviation $\delta H_0 = G_z z$ the overlapping modes provide a resolvable structure. The small inhomogeneity created by epoxy may be neglected in this case. A negative $G_z$ implies that the
potential shape is a box with tilted bottom decreasing downwards as $U(z)=\gamma_{e} G_{z} z$ (insert in Fig.~\ref{NegGradSpectra}). The oscillating
solutions of Eq.~(\ref{ISR equation}) are found in the form of Bessel functions in the radial and Airy functions in the axial direction. The
shape of the solution should match the shape of the rf. field excitation, which is axially symmetric and has a maximum at $r=0$. Since only the
zeroth-order Bessel function satisfies this condition, the solution is written as
\begin{equation}
S_{+}\sim J_{0}(\kappa r)e^{i\omega t}Ai(z/\lambda+\lambda ^2 k_{z}^2),
\label{Solution}
\end{equation}
with $\kappa^2=\mu \omega/D_0$ and $\lambda=(D_0/\gamma_{e}G_{z}\mu)^{1/3}$ being the length scale of the $Ai(z)$ function. We apply reflective
boundary conditions with zero spin flux through the walls of the cylinder. These conditions define the eigenvalues of $\kappa$ and $k_z$,
eigenfrequencies $\omega$ of the oscillations, and the number of $Ai$ function oscillations in the axial direction. For $n=5\times10^{17}$
cm$^{-3}$ and $T=260$ mK we estimate $\lambda\approx7\,\mu$m. In order to generate spin waves, their wave function given by Eq.~(\ref{Solution}) should
overlap with the rf. field. This implies that in a negative gradient we may excite only high-lying modes, which have their frequency eigenvalues in the vicinity of the intersection of the linear potential and the upper wall (see insert in Fig.~\ref{NegGradSpectra}). The total number of modes counting from the bottom of the potential U(z) and the corresponding number of $Ai(z)$ oscillations are $\gg L/ \lambda \sim 100$ even for our shortest samples. Therefore, we deal with a quasi-continuum of mode frequencies, and excite only those with non-vanishing wave
functions near the top of the cylinder. The position of the lower boundary does not influence the shape of these modes near the top because of
the rapid oscillation of $Ai(z)$ near the bottom. This explains why the observed ESR lineshapes do not depend on the sample height.
\begin{figure}
\includegraphics[width=8 cm]{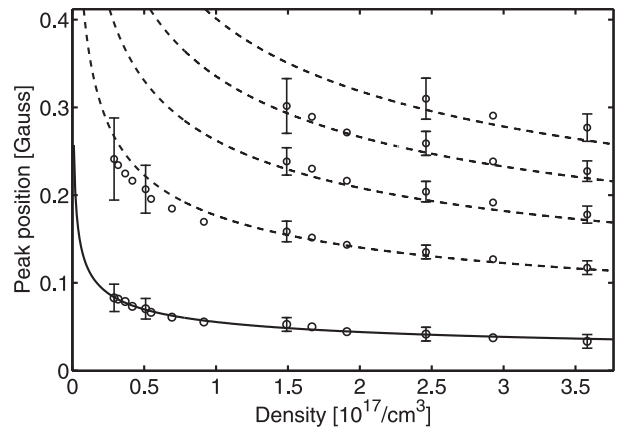}
\caption{Positions of magnon peaks for samples in large negative magnetic field gradients of -4 G/mm as functions of H gas density.}
\label{PeakPos}
\end{figure}

Since the position of the lower boundary does not influence the shape of $S_{+}$ near the top, we consider the case of a semi-infinite cylinder. In this case the spin waves may freely propagate downwards along the slope of the magnetic potential. Sweeping the magnetic field $h$ we scan through the region $l_{EF}$ by tuning the rf. field in resonance with electron spins. At each point the transversal magnetization is created
locally and two spin waves are launched; one up and the other down. For zero axial gradient these would be similar to the
electromagnetic waves in a cylindrical waveguide. Moving the point of excitation with respect to the top we scan a standing wave pattern
formed by the direct and reflected waves with the spatial period equal to half-wavelength. For non-zero gradients the sine-waves are accelerated
downwards and are replaced by the Airy functions $Ai(z/\lambda)$. The contribution of spin wave modes to the ESR absorption is then $\sim F(h) =
Ai^{2}(h/(G_{z}\lambda)) \times H_{1}^{2}(h/G_z)$, shown by the dotted line in Fig.~\ref{NegGradSpectra}. This will produce maxima of absorption
at the antinodes of $Ai(z/\lambda)$, the points $z_i$ corresponding to the roots $x_i$ of $dAi(x)/dx$. The peak positions in the field sweep are $h_i=G_z \lambda x_i +B_0$. They have a $n^{-1/3}$ density dependency since the diffusion coefficient is inversely proportional to the H density; $n D_0=1.5\times 10^{18}$ cm$^{-3}$ \cite{Lhuillier83}. The separation between the peaks is proportional to the roots of the $Ai$ function,
which we verified as follows. First we fit the position of the leftmost peak by the relation $h_1 = C n^{-1/3} x_1 + B_0$ with $C$
and $B_0$ being fitting parameters. Then we calculate the positions of the next four peaks as functions of density as $h_i = C n^{-1/3} x_i +
B_0$. The results plotted by dashed lines in Fig.~\ref{PeakPos} match the experimental data within the error bars. The relation $C=(G_{z}^2 n D_0/ \gamma_e \mu)^{1/3}$ provides the measurement of the ratio $n D_0 / \mu=1.9(4)\times10^{17}$ cm$^{-3}$. From
the fit we find $\mu=8(2)$, which has been confirmed by numerical solutions of Eq.~(\ref{ISR equation}). The value of $\mu$ obtained in
our work is in good agreement with the results for nuclear spin waves in H$\downarrow$ \cite{Johnson84}.

The feature (ii) was observed for small or zero linear gradients. In this case there is a local magnetic field maximum near the top of the sample and close to EF. An example of a $\delta B(r,z)$ profile calculated for $G_z =0$ is
presented in Fig.~\ref{cell} (b). Together with the wall of the tube such a field profile forms a 3D trap for the electron spin waves. A numerical
solution of Eq.~(\ref{ISR equation}) reveals a large number of modes in this trap with the separation of a few kHz, again too small to be
resolved. Sweeping the static field offset $h$ we tune our excitation to resonance in different regions of the sample volume. Absorption is
increased as energy is pumped into the spin wave modes and the absorption strength depends on the density of modes for a given frequency
and spatial position. Applying a small linear gradient we modify the axial profile and consequently move the trap bottom up or down as shown in
Fig.~\ref{cell} (b). The position of the corresponding spin wave peak in the ESR spectrum is determined by the static field at the trap
bottom, which leads to a nonlinear dependence of the peak position on $G_z$ (Fig.~\ref{Spectra in Grad} insert). This also explains why we cannot see the feature (ii) on the left side of the main ESR line. For large positive gradients the field maximum is always located at the top of the
cylinder, coinciding with the position of the main ESR line at the $H_{1}^{2}(z)$ maximum. Note that if the peak (ii) would be caused by
the second rf. field maximum located some what below the main one, changing the sign of the field gradient would mirror the
spectrum and this peak (ii) would be on the left hand side with linear dependence on $G_z$. Remarkably, the spin-wave
modes (ii) are well excited and detected even by the very weak tail of the rf. field. The  spin wave signal is much larger than absorption due to the ESR itself. Comparing the density dependence of the two spin wave we find that the spin waves (ii)
survive at much smaller gas densities than the propagating modes (i). These observations confirm that in case (i) we deal with a spin wave-guide and with a 3D spin wave trap in case (ii).

We have also considered the possibility of spin excitations called
magnetostatic Walker modes \cite{Walker} which should exhibit a strong dependence of the mode positions on the sample geometry. However, this was not observed in our experiments. The magnetostatic modes do not depend on temperature, because they are caused by long range dipolar
forces and do not involve atomic collisions. We verified that the separation between peaks (i) increases slightly at higher temperatures. For
the ISR modes we expect a temperature dependence $D_0 / \mu \sim\sqrt{T}$ or weaker \citep{Lhuillier83}, which is in line with our observations.
We conclude that both types of the spin wave modes described above are caused by the ISR effect.

In this work we have studied electron spin waves in atomic hydrogen quantum gas. We detected two types of spin wave excitations; travelling
modes guided by the cylindrical spin wave-guide and modes confined in the magnetic potential well. Similarly to the matter wave interference of
cold atoms \cite{Bloch2000, Harkonen06} the presence of a reflective boundary in the spin-waveguide leads to interference of the two travelling
spin waves, one falling straight down, and the other one reflecting from the upper wall. In the quantum regime the spin waves are described as
quasiparticles called magnons \cite{Bunkov07, Demokritov06}. Our trapping technique resembles the original technique for trapping cold gases,
but here it is realized for high field seeking quasiparticles. In further experiments we hope to observe effects of statistical correlations
between trapped magnons leading to Bose-Einstein condensation and spin superfluidity.

\begin{acknowledgments}
This work was supported by the Academy of Finland (Grants No. 122595, and 133682) and the Wihuri Foundation. We thank D. M. Lee, N. Bigelow, G. Volovik, V. Eltsov, S. Jaakkola, J.T.M. Walraven, and L. Lehtonen for valuable discussions.
\end{acknowledgments}

\bibliography{Magnons}

\end{document}